\documentclass[twoside,11pt]{9ssmmp} %%% Please do not change this class !!! %%%
\usepackage{fancyhdr}
\usepackage{amsfonts}
\usepackage{amssymb}
\usepackage{graphicx}
\usepackage{hyperref}

\newcommand\rf[1]{(\ref{eq:#1})}
\newcommand\lab[1]{\label{eq:#1}}
\newcommand\nonu{\nonumber}
\newcommand\br{\begin{eqnarray}}
\newcommand\er{\end{eqnarray}}
\newcommand\be{\begin{equation}}
\newcommand\ee{\end{equation}}

\newcommand\lb{\lbrack}
\newcommand\rb{\rbrack}

\newcommand\llb{\left\lbrack}
\newcommand\rrb{\right\rbrack}

\renewcommand\({\left(}
\renewcommand\){\right)}
\newcommand\bv{\bigm\vert}               %%
\newcommand\bgv{\bigg\vert}              %%

\newcommand\bc{\begin{center}}
\newcommand\ec{\end{center}}

\newcommand\partder[2]{\frac{{\partial {#1}}}{{\partial {#2}}}}
\renewcommand\a{\alpha}

\renewcommand\d{\delta}
\newcommand\eps{\epsilon}
\newcommand\vareps{\varepsilon}

\newcommand\G{\Gamma}

\newcommand\h{\frac{1}{2}}
\renewcommand\k{\kappa}
\renewcommand\l{\lambda}

\newcommand\m{\mu}
\newcommand\n{\nu}

\renewcommand\O{\Omega}

\newcommand\vp{\varphi}
\renewcommand\P{\Phi}
\newcommand\pa{\partial}

\renewcommand\th{\theta}

\newcommand\wti{\widetilde}
\newcommand\cA{{\mathcal A}}

\newcommand\cC{{\mathcal C}}

\newcommand\cU{{\mathcal U}}

\newcommand{\ct}[1]{\cite{#1}}
\newcommand{\bib}[1]{\bibitem{#1}}

\newcommand\PRL[3]{\textsl{Phys. Rev. Lett.} \textbf{#1} (#2) #3}

\newcommand\PRD[3]{\textsl{Phys. Rev.} \textbf{D#1} (#2) #3}

\newcommand\PLB[3]{\textsl{Phys. Lett.} \textbf{#1B} (#2) #3}

\newcommand\AoP[3]{\textsl{Ann. of Phys.} \textbf{#1} (#2) #3}

\newcommand\IJMPA[3]{\textsl{Int. J. Mod. Phys.} \textbf{A#1} (#2) #3}
\newcommand\IJMPD[3]{\textsl{Int. J. Mod. Phys.} \textbf{D#1} (#2) #3}

\newcommand\MPLA[3]{\textsl{Mod. Phys. Lett.} \textbf{A#1} (#2) #3}

\newcommand\vpdot{\stackrel{.}{\varphi}}

\newcommand\adot{\stackrel{.}{a}}

\newcommand\utdot{\stackrel{.}{\widetilde u}}
\newcommand\Adot{\stackrel{.}{A}}

\begin{document}

%% Please modify the following line to include the title of your contribution 
%% and acknowledgments:

\title{Quintessence, Unified Dark Energy and Dark Matter, and Confinement/Deconfinement Mechanism
% \hspace{.25mm}\thanks{\,This work has been supported by\dots }
}

%% Please modify the following lines to include author names, affiliations 
%% and e-mail addresses:

\author{
\bf{Eduardo Guendelman}\hspace{.25mm}\thanks{\,e-mail address:
guendel@bgu.ac.il}
\\ \normalsize{Physics Department, Ben Gurion University of the Negev}\\
\normalsize{Beer Sheva, Israel} \vspace{2mm} \\ 
\bf{Emil Nissimov and Svetlana Pacheva}\hspace{.25mm}\thanks{\,e-mail
address: nissimov@inrne.bas.bg, svetlana@inrne.bas.bg}
\\ \normalsize{Institute for Nuclear Research and Nuclear Energy}\\
\normalsize{Bulgarian Academy of Sciences, Sofia, Bulgaria} }
% \vspace{2mm} \\ \bf{Third author}\hspace{.25mm}\thanks{\,e-mail
% address: third@4ss.mmp}
% \\ \normalsize{Address of the third author}

\date{} %% Please do not modify

\maketitle %% Please do not modify

\begin{abstract}
% Please add the text of the abstract here. 
% Here you can also add relevant PACS and MSC classification codes.
We describe a new type of generalized gravity-matter models where 
$f(R)=R+R^2$ gravity couples in a non-conventional way to a scalar
``inflaton'' field, to a second scalar ``darkon'' field responsible for dark
energy/dark matter unification,  as well as to a non-standard nonlinear gauge 
field system containing a square-root of the ordinary Maxwell Lagrangian,
which is responsible for a charge confining/deconfinfing mechanism. 
% The latter is known to describe charge confinement in flat spacetime. 
The essential non-conventional feature of our models is employing the formalism of 
non-Riemannian volume forms, \textsl{i.e.} metric-independent 
non-Riemannian volume elements on the spacetime manifold, defined in terms of 
auxiliary antisymmetric tensor gauge fields. 
Although being (almost) pure-gauge degrees of freedom, 
the non-Riemannian volume-forms trigger a series of important features 
unavailable in ordinary gravity-matter models. Upon passing to the physical
Einstein frame we obtain an effective matter-gauge-field Lagrangian of
quadratic ``k-essence'' type both w.r.t. the ``inflaton'' and the ``darkon'', 
with the following properties: 
(i) Remarkable effective ``inflaton'' 
potential possessing two infinitely large flat regions with vastly different
heights (``vacuum'' energy densities) describing the ``early'' and ``late'' 
Universe; 
(ii) Nontrivial effective gauge coupling constants running with the
``inflaton'', in particular, effective ``inflaton''-running coupling
constant of the square-root Maxwell term, which determines the strength of
the charge confienement; 
(iii) The confinement-strength gauge coupling constant is non-zero in the
``late'' Universe, \textsl{i.e.}, charge confinement is operating, 
whereas it vanishes in the ``early'' Universe, \textsl{i.e.},
confinement-free epoch;
(iv) The unification of dark energy and dark matter is
explicitly seen within the FLRW reduction, where they appear as dynamically 
generated effective vacuum energy density and dynamically induced dust-like 
matter, correspondingly.
%%%%%%%%%%%%%%%%%%%%%%%%
% PACS numbers: 04.50.Kd, % Modified theories of gravity
% 98.80.Bp, % Origin and formation of the Universe
% 95.36.+x % Dark energy
% 12.38.Aw % Quark confinement
%%%%%%%%%%%%%%%%%%%%%%%%
\end{abstract}

%%%%%%%%%%%%%%%%%%%%%%%%%%%%%%%%%%%%%%%%%%%%%%%%%%%%%%%%%%%%%%%%%%%%%%%%%%%%%%%%%%
\section{Introduction}
\label{intro}

Our main task, which pertains to the interface of particle physics and cosmology 
\ct{general-cit}, is to present a non-canonical model of extended ($f(R)=R+R^2$)
\ct{f(R)-gravity}
gravity interacting in a non-standard way with two scalar fields and a
strongly nonlinear gauge field, which is capable to provide a systematic
description of:

(a) Unified treatment of dark energy and dark matter (for a background, see
\ct{dark-energy-observ,DM-rev}) revealing them as manifestations of a
single material entity -- the first non-canonical scalar ``darkon'' field. A
number of proposals already exist for an adequate description of dark energy's 
and dark matter's dynamics within the framework of standard general relativity 
or its modern extensions, among them: ``Chaplygin gas'' models \ct{chaplygin}, 
``purely kinetic k-essence'' models \ct{purely-kinetic-k-essence}, ``mimetic''
dark matter models \ct{mimetic}. In Section 2 we briefly review 
our own approach \ct{dusty-1-dusty-2}.

(b) Quintessential scenario driven by the remarkable dynamically generated
effective potential of the second ``inflaton'' scalar field with a consistent 
explanation of the vast difference between the energy scales of the ``early'' 
and the ``late'' Universe;

(c) Charge confinement/deconfinement mechanism triggered via special interplay 
between the ``inflaton'' and the nonlinear gauge field dynamics explaining
absence of charge confinement in the ``early'' Universe and exhibiting 
confinement in the ``late'' Universe.

%%%%%%%%%%%%%%%
The principal ingredient of our approach is the method of non-Riemannian
volume-forms (metric-independent volume elements) on the pertinent spacetime
manifold (see refs.\ct{susyssb-1,belgrade-14,grav-bags} for a consistent 
geometrical formulation, which is an extension of the originally proposed method
\ct{TMT-orig}). Non-Riemannian volume-forms are constructed in terms of auxiliary
maximal-rank antisymmetric tensor gauge fields, which are shown to be 
essentially pure-gauge degrees of freedom, \textsl{i.e.}, they do not 
introduce additional propagating field-theoretic degrees of freedom. Yet,
they leave a trace in a form of several dynamically induced integration
constants, which trigger a series of important features unavailable in ordinary 
gravity-matter models. 

Another important ingredient of our approach is the inclusion of interaction
of the extended gravity with a nonlinear gauge field system containing
alongside the ordinary Maxwell Lagrangian also a square-root of the latter.
In flat spacetime such gauge field system is known \ct{GG-1,GG-2} to yield a simple
implementation of  `t Hooft's idea \ct{thooft} about confinement being produced
due to the presence in the energy density of electrostatic field  configurations 
of a term {\em linear} w.r.t. electric displacement field in the infrared
region (presumably as an appropriate infrared counterterm). It has been
shown in \ct{GG-1} (for flat spacetime) and in appendix B of \ct{grav-bags}
(in curved spacetime) that the strength of confinement in this model is 
measured by the corresponding coupling constant of the square-root Maxwell term.
Let us also note that one could start as well with the non-Abelian version
of the above nonlinear gauge theory containing square-root of the Yang-Mills
Lagrangian. For static spherically symmetric solutions, the non-Abelian theory 
effectively reduces to an Abelian one as pointed out in \ct{GG-1}.
% Thus, we will consider below the dynamics of the Abelian nonlinear gauge
% system as mimickimg quantum chromodynamics

In Section 3 we obtain upon passing to the physical
Einstein frame an effective matter-gauge-field Lagrangian of
quadratic ``k-essence'' type \ct{k-essence-orig} w.r.t. both the ``inflaton''
and the ``darkon'' scalar fields with several remarkable properties: 

(i) It contains an effective ``inflaton'' potential possessing two infinitely 
large flat regions with vastly different heights -- dynamically generated 
``vacuum'' energy densities thanks to the appearance of the above mentioned
integration constants --  which describe the ``early'' and ``late'' 
Universe, accordingly; 

(ii) The Einstein-frame Lagrangian contains nontrivial effective gauge 
coupling constants running with the ``inflaton''. Particularly important is
the effective ``inflaton''-running coupling constant of the square-root Maxwell 
term, which determines the ``inflaton''-dependent strength of the charge 
confienement; 

(iii) We show the confinement-strength gauge coupling constant to be non-zero 
in the ``late'' Universe, \textsl{i.e.}, charge confinement is operating there. 
On the other hand, the confinement-strength gauge coupling constant vanishes 
in the ``early'' Universe, \textsl{i.e.}, the latter being confinement-free epoch;

In Section 4 we consider a cosmological FLRW reduction of the Einstein-frame 
effective matter-gauge-field Lagrangian where the unification of dark energy 
and dark matter is explicitly seen upon identifying them  as dynamically 
generated effective vacuum energy density and dynamically induced dust-like 
matter, correspondingly.

%%%%%%%%%%%%%%%%%%%%%%%%%%%%%%%%%%%%%%%%%%%%%%%%%%%%%%%%%%%%%%%%%%%%%%%%%%%%%%%%%%
\section{A Simple Model of Dark Energy and Dark Matter Unification}
\label{dusty}

We start with a simple particular case of a non-conventional 
gravity-scalar-field action -- a member of the general class of the 
non-Riemannian-volume-element-based gravity-matter theories \ct{emergent,grav-bags}
(we use for simplicity units where the Newton constant $G_N = 1/16\pi$):
\be 
S = \int d^4 x \sqrt{-g}\, R +
\int d^4 x \bigl(\sqrt{-g}+\P(C)\bigr) L(u,Y) \; .
\lab{TMT-0}
\ee
$R$ denotes the standard Riemannian scalar curvature for the spacetime
Riemannian metric $g_{\m\n}$. 
$L(u,Y)$ is general-coordinate invariant Lagrangian of a single scalar field 
$u (x)$, the simplest example being:
\be
L(u,Y) = Y - V(u) \quad ,\quad Y \equiv - \h g^{\m\n}\pa_\m u \pa_\n u \; ,
\lab{standard-L}
\ee
It is coupled symmetrically to two mutually independent spacetime volume-elements 
(integration measure densities) -- the standard Riemannian $\sqrt{-g}$ and to 
an alternative  non-Riemannian (metric-independent) one:
\be
\P(C) = \frac{1}{3!}\vareps^{\m\n\k\l} \pa_\m C_{\n\k\l} \; .
\lab{mod-measure}
\ee
As a result of the equations of motion w.r.t. $C_{\m\n\l}$ we obtain
a crucial new property -- {\em dynamical constraint} on $L(u,Y)$:
\be
\pa_\m L (u,Y) = 0 \; \longrightarrow \; L(u,Y) = - 2M_0 = {\rm const} \; ,
 \; \mathrm{i.e.} \; Y = V(u) - 2M_0 \; .
\lab{L-const}
\ee
% \textsl{i.e.}, $Y = V(u) - 2M_0$.
The integration constant $M_0$ will play below (Eq.\rf{T-hydro}) the role of 
dynamically generated cosmological constant.

The pertinent energy-momentum tensor $T_{\m\n}$ % and $J^\m$ \rf{J-conserv}
can be cast into a relativistic hydrodynamical form 
(taking into account \rf{L-const}):
\br
T_{\m\n} = - 2M_0 g_{\m\n} + \rho_0 u_\m u_\n \; ,
\lab{T-hydro} \\
\rho_0 \equiv \Bigl(1+\frac{\P(C)}{\sqrt{-g}}\Bigr)\, 2Y % \partder{L}{Y} 
\;\;\; ,\;\; 
u_\m \equiv - \frac{\pa_\m u}{\sqrt{2Y}} \;\; ,\;\; 
 u^\m u_\m = -1 \; ,
\lab{rho-0-def}
\er
with the corresponding pressure $p$ and energy density $\rho$:
\be
p = - 2M_0 = {\rm const} \quad ,\quad
\rho =  \rho_0 - p = 2M_0 + \Bigl(1+\frac{\P(C)}{\sqrt{-g}}\Bigr)\, 2Y \; .
\lab{p-rho-def}
\ee

Because of the constant pressure ($p=-2M_0$) the covariant energy-momentum
conservation $\nabla^\n T_{\m\n}=0$ implies {\em both} conservation 
of a hidden Noether symmetry current $J^\m = \rho_0 u^\m$, 
as well as {\em geodesic fluid motion}:
\be
\nabla_\m J^\m  \equiv \nabla_\m \bigl(\rho_0 u^\m\bigr) = 0 \quad ,\quad 
u_\n \nabla^\n u_\m = 0 \; .
\lab{dust-geo}
\ee
The {\em hidden strongly nonlinear Noether symmetry} giving rise to the
current $J^\m = \rho_0 u^\m$ results from the invariance (up to a total 
derivative) of the scalar field action in \rf{TMT-0}, exclusively due to 
the presence of the non-Riemannian volume element $\P (C)$, under the 
following nonlinear symmetry transformations:
\br
\d_\eps u = \eps \sqrt{Y} \quad ,\quad \d_\eps g_{\m\n} = 0 \quad ,\quad
% \nonu \\
% \d_\eps C_{\m\n\l} = - \eps \frac{1}{2\sqrt{X}} \vareps_{\m\n\l\k}
% g^{\k\r}\pa_\r u \, \bigl(\P(C) + \sqrt{-g}\bigr)  \; .
\d_\eps \cC^\m = - \eps \frac{1}{2\sqrt{Y}} g^{\m\n}\pa_\n u 
\bigl(\P(C) + \sqrt{-g}\bigr)  \; ,
\lab{hidden-sym}
% \nonu
\er
where $\cC^\m \equiv \frac{1}{3!} \vareps^{\m\n\k\l} C_{\n\k\l}$.

Thus, $T_{\m\n}$ \rf{T-hydro}-\rf{p-rho-def} represents an exact sum of two
contributions of the two ``dark'' material species with corresponding
pressures and energy densities:
\br
p = p_{\rm DE} + p_{\rm DM} \quad,\quad \rho = \rho_{\rm DE} + \rho_{\rm DM}
\lab{DE+DM-1} \\
p_{\rm DE} = -2M_0\;\; ,\;\; \rho_{\rm DE} = 2M_0 \quad ; \quad
p_{\rm DM} = 0\;\; ,\;\; \rho_{\rm DM} = \rho_0 \; ,
\lab{DE+DM-2}
\er
\textsl{i.e.}, according to \rf{dust-geo} the dark matter component is a 
dust fluid flowing along geodesics. 
This is explicit unification of dark energy and dark matter
originating from the non-canonical dynamics of a single scalar field --
the ``darkon'' $u(x)$ coupled symmetrically to a standard Riemannian and another 
non-Riemanian (metric-independent) volume element. Let us also note that the
physical result $T_{\m\n}$ \rf{T-hydro} does {\em not} depend on the
explicit form of the ``darkon'' potential $V(u)$.

Upon reduction to the cosmological class of
Friedmann-Lemaitre-Robertson-Walker (FLRW) metrics:
\be
ds^2 = - dt^2 + a^2(t) \Bigl\lb \frac{dr^2}{1-K r^2}
+ r^2 (d\th^2 + \sin^2\th d\phi^2)\Bigr\rb \; ,
\lab{FLRW-0}
\ee
the $J^\m$-current conservation \rf{dust-geo} now reads:
\br
\nabla^\m \bigl(\rho_0 u_\m\bigr) = 0 \;\; \to \;\; 
\frac{d}{dt}\bigl(a^3 \rho_0\bigr) = 0  \;\; \to \;\;
% \rho_0 \equiv \Bigl(1+\frac{\P(B)}{a^3}\Bigr)\, 2X \partder{L}{X} = 
\rho_0 = \frac{c_0}{a^3} \; \;, \;\; c_0 = {\rm const},
\lab{dust-sol} \\
p = - 2M_0 \quad ,\quad \rho = 2M_0 + \frac{c_0}{a^3} \; ,
\lab{p-rho-FLRW}
\er
with the dark matter contribution $\rho_0$ being the typical cosmological 
dust solution (\textsl{e.g.} \ct{Lambda-CDM-2}).

%%%%%%%%%%%%%%%%%%%%%%%%%%%%%%%%%%%%%%%%%%%%%%%%%%%%%%%%%%%%%%%%%%%%%%%%%%%%%%%%%%
\section{Noncanonical Gravity-Matter System Coupled to Charge-Confining 
Nonlinear Gauge Field}
\label{buggy-base}

Let us now extend the simple gravity-``darkon'' model \rf{TMT-0} to 
$f(R)=R+R^2$ gravity coupled non-canonically to 
both ``inflaton'' $\vp (x)$ and ``darkon'' $u(x)$ scalar fields within the
non-Riemannian volume-form formalism, as well as we will also add 
coupling to a non-standard non-linear gauge field subsystem:
\br
S = \int d^4 x\,\P (A) \Bigl\lb g^{\m\n} R_{\m\n}(\G) + X - V_1 (\vp)
 -\h f_0 \sqrt{-F^2}\Bigr\rb +
\nonu \\
\int d^4 x\,\P (B) \Bigl\lb \eps R^2 - \frac{1}{4e^2}F^2
+ \frac{\P (H)}{\sqrt{-g}}\Bigr\rb
% \nonu \\
+ \int d^4 x \bigl(\sqrt{-g}+\P(C)\bigr) L(u,Y) \; .
% \phantom{aaaaaaaaaa}
\lab{TMMT-1}
\er
Here the following notations are used:

\begin{itemize}
% \item
% $L(u,Y)$ is the same as in \rf{TMT-0}-\rf{standard-L}; 
% $X \equiv -\h g^{\m\n} \pa_\m \vp \pa_\n \vp$.
\item
$\P(A)$ and $\P(B)$ are two new metric-independent non-Riemannian
volume-elements:
\be
\P (A) = \frac{1}{3!}\vareps^{\m\n\k\l} \pa_\m A_{\n\k\l} \quad ,\quad
\P (B) = \frac{1}{3!}\vareps^{\m\n\k\l} \pa_\m B_{\n\k\l} \; ,
\lab{Phi-1-2}
\ee
defined in terms of field-strengths of two auxiliary 3-index antisymmetric
tensor gauge fields, apart from $\P (C)$; 
\item
$\P (H) = \frac{1}{3!}\vareps^{\m\n\k\l} \pa_\m H_{\n\k\l}$  
% \lab{Phi-H}
% \er
is the dual field-strength of an additional auxiliary tensor gauge field 
$H_{\n\k\l}$ whose presence is crucial for the consistency of \rf{TMMT-1}.
\item
Let us specifically emphasize the importance to use here the Palatini formalism:  
$R=g^{\m\n} R_{\m\n}(\G)$ ; $g_{\m\n}$, $\G^\l_{\m\n}$ -- metric and affine
connection are {\em apriori} independent.
\item
The ``inflaton'' part reads:
\be
X \equiv -\h g^{\m\n} \pa_\m \vp \pa_\n \vp \quad , \quad
V_1(\vp)= f_1 \exp\{-\a\vp\} \; .
\lab{V-1-def}
\ee
\item
$F_{\m\n}$ is the field-strength of an (Abelian) gauge field $\cA_{\m}$:
\be
F_{\m\n} = \pa_\m \cA_{\n} - \pa_\n \cA_{\m} \quad ,\quad
F^2 = F_{\m\n} F_{\k\l} g^{\m\k} g^{\n\l}
\lab{F-def}
\ee
\item
The coupling parameters $f_1$, $f_0$ and $\a$ are dimensionful positive
constants (mass$^4$, mass$^2$, mass$^{-1}$).
\end{itemize}

The specific form of the action \rf{TMMT-1} apart from the ``darkon'' field
part \rf{TMT-0} has been fixed by the requirement for invariance under 
global Weyl-scale transformations:
\br
g_{\m\n} \to \l g_{\m\n} \;,\;\G^\m_{\n\l} \to \G^\m_{\n\l}
\;,\; \vp \to \vp + \frac{1}{\a}\ln \l \;,\; \cA_\m \to \cA_\m \; ,
\nonu \\
A_{\m\n\k} \to \l A_{\m\n\k} \;\; ,\;\; 
B_{\m\n\k} \to \l^2 B_{\m\n\k} \;\; ,\;\; H_{\m\n\k} \to H_{\m\n\k} \; .
\lab{scale-transf}
\er

As shown in \ct{grav-bags}, the solution to the equations of motion w.r.t.
independent affine connection $\G^\m_{\n\l}$ yield the latter as a
Levi-Civitta connection:
\be
\G^\m_{\n\l} = \G^\m_{\n\l}({\bar g}) = 
\h {\bar g}^{\m\k}\(\pa_\n {\bar g}_{\l\k} + \pa_\l {\bar g}_{\n\k} 
- \pa_\k {\bar g}_{\n\l}\) \; ,
\lab{G-eq}
\ee
w.r.t. the Weyl-rescaled metric:
\be
{\bar g}_{\m\n} = (\chi_1 + 2\eps \chi_2 R) g_{\m\n} \;\; ,\;\; 
\chi_1 \equiv \frac{\P_1 (A)}{\sqrt{-g}} \;\; ,\;\;
\chi_2 \equiv \frac{\P_2 (B)}{\sqrt{-g}} \; .
\lab{bar-g}
\ee

Varying the action \rf{TMMT-1} w.r.t. auxiliary tensor gauge fields
$A_{\m\n\l}$, $B_{\m\n\l}$ and $H_{\m\n\l}$ yields the equations:
\br
\pa_\m \Bigl\lb R + X - V_1 (\vp) -\h f_0 \sqrt{-F^2}\Bigr\rb = 0 
\nonu \\
\pa_\m \Bigl\lb \eps R^2 - \frac{1}{4e^2}F^2 + \frac{\P (H)}{\sqrt{-g}}\Bigr\rb = 0 
\quad, \quad \pa_\m \Bigl(\frac{\P_2 (B)}{\sqrt{-g}}\Bigr) = 0 \; ,
\lab{A-B-H-eqs}
\er
whose solutions are:
\br
R + X - V_1 (\vp) -\h f_0 \sqrt{-F^2} = M_1 = {\rm const} \;,
\nonu \\
\eps R^2 - \frac{1}{4e^2}F^2 + \frac{\P (H)}{\sqrt{-g}} = - M_2  = {\rm const} 
\;\; , \;\;
\frac{\P_2 (B)}{\sqrt{-g}} \equiv \chi_2 = {\rm const} \;.
\lab{integr-const}
\er
Here $M_1$ and $M_2$ are arbitrary dimensionful (mass$^4$) and $\chi_2$
arbitrary dimensionless integration constants. 
The appearance of $M_1,\, M_2$ signifies {\em dynamical spontaneous
breakdown} of global Weyl-scale invariance under \rf{scale-transf} due to the 
scale non-invariant solutions (second and third ones) in \rf{integr-const}.

The physical meaning of the integration constants $M_0,\,M_{1,2},\,\chi_2$,
has been elucidated in refs.\ct{belgrade-14,susyssb-2,grav-bags} from 
the point of view of the canonical Hamiltonian formalism. 
Namely, it was shown that $M_0,\,M_{1,2},\,\chi_2$,
which remain the only traces of the auxiliary gauge fields
$C_{\m\n\l},\,A_{\m\n\l},$ $\,B_{\m\n\l},\,H_{\m\n\l}$, 
are identified as conserved Dirac-constrained canonical momenta conjugated to 
(certain components of) the latter.

Using the equations of motion of the starting action \rf{TMMT-1} 
w.r.t. $g_{\m\n}$ as well as Eqs.\rf{integr-const} we obtain the following
relation between the original metric $g_{\m\n}$ and the Weyl-rescaled one 
\rf{bar-g} (with $\chi_{1,2}$ the same as in \rf{bar-g}, $Y$ - same as in
\rf{standard-L}):
\br
{\bar g}_{\m\n} = \chi_1 \O g_{\m\n} \quad ,\quad
\chi_1 \O = \frac{\chi_1 + 2\eps\chi_2\bigl(V_1(\vp)+M_1\bigr)}{1 +
2\eps\chi_2\bigl({\bar X} - \frac{f_0}{2}\sqrt{-{\bar F}^2}\bigr)} \; ,
\lab{bar-g-eq} \\
\chi_1 = \bigl(V_1(\vp)+M_1\bigr)^{-1}
\Bigl\lb 2\chi_2 M_2 - 4M_0 - (1+\chi_4)Y\Bigr\rb \; ,
\lab{chi-1-eq}
\er
where:
\be
{\bar X} \equiv -\h g^{\m\n} \pa_\m \vp \pa_\n \vp \;\; , \;\;
{\bar F}^2 \equiv F_{\m\n} F_{\k\l} {\bar g}^{\m\k} {\bar g}^{\n\l} 
\;\; , \;\; \chi_4 \equiv \frac{\P(C)}{\sqrt{-g}}
\lab{bar-def}
\ee

Now, following the same steps as in refs.\ct{grav-bags} we derive from \rf{TMMT-1} 
the physical {\em Einstein-frame} theory w.r.t. Weyl-rescaled Einstein-frame
metric ${\bar g}_{\m\n}$ \rf{bar-g-eq} and perform an
additional ``darkon'' field redefinition $u \to {\wti u}$: 
\be
\;\; \partder{{\wti u}}{u}=\bigl( V_1 (u) - 2M_0\bigr)^{-\h} \quad ; \quad
Y \to {\wti Y} = - \h {\bar g}^{\m\n}\pa_\m {\wti u} \pa_\n {\wti u} \; .
\lab{darkon-redef}
\ee
The explicit form of the Einstein-frame matter action reads:
\br
L_{\rm eff} = {\bar X} - \frac{f_0}{2} \sqrt{-{\bar F}^2} -
\frac{\chi_2}{e^2} {\bar F}^2 
+ \eps\chi_2 \Bigl({\bar X} - \frac{f_0}{2}\sqrt{-{\bar F}^2}\Bigr)^2 
\nonu \\
- {\wti Y} \Bigl(V(\vp) + M_1\Bigr)\Bigl\lb 1 + 
2 \eps\chi_2 \Bigl({\bar X} - \frac{f_0}{2}\sqrt{-{\bar F}^2}\Bigr)\Bigr\rb
\nonu \\
+ {\wti Y}^2 \left\{\chi_2\Bigl\lb M_2 + \eps\bigl(V(\vp) + M_1\bigr)^2\Bigr\rb 
- 2M_0 \right\}
\; .
\lab{L-eff}
\er

To reveal the total matter effective potential $U_{\rm total}$, \textsl{i.e.}, the
{\em minus} part of $L_{\rm eff}$ \rf{L-eff} for static (spacetime independent) 
matter fields $\vp$, $F^2$ and $u$ ({\em not} ${\wti u}$ \rf{darkon-redef}),
we note that relations \rf{L-const}, \rf{bar-g-eq} and \rf{darkon-redef} imply 
${\wti Y} = \frac{1}{\chi_1 \O}$, meaning that according to 
\rf{bar-g-eq}-\rf{chi-1-eq} for static matter fields:
\be
{\wti Y}\bgv_{\rm static} = \frac{\bigl(V(\vp) + M_1\bigr)\bigl( 1 - 
\eps\chi_2 f_0 \sqrt{-{\bar F}^2}\bigr)}{2
\left\{\chi_2\Bigl\lb M_2 + \eps\bigl(V(\vp) + M_1\bigr)^2\Bigr\rb - 2M_0\right\}}
\; .
\lab{Y-tilde-static}
\ee
Upon inserting \rf{Y-tilde-static} in \rf{L-eff} we find:
\br
U_{\rm total}\bigl(\vp, {\bar F}^2\bigr) = 
- L_{\rm eff}\bgv_{\rm static} = 
\cU (\vp) + \h f_{\rm eff}(\vp) \sqrt{-{\bar F}^2} + 
\frac{1}{4e^2_{\rm eff}(\vp)} {\bar F}^2 \; ,
\lab{U-total} \\
\cU (\vp)\equiv \frac{\bigl(V_1(\vp) + M_1\bigr)^2}{4
\left\{\chi_2\Bigl\lb M_2 + \eps\bigl(V_1(\vp) + M_1\bigr)^2\Bigr\rb - 2M_0\right\}}
\;,
\lab{cU-def} \\
\bigl( \mathrm{recall ~\rf{V-1-def}}\;\; V_1(\vp) = f_1 \exp\{-\a\vp\} \bigr)\; ,
\phantom{aaaaaa}
\nonu
\er
% (recall $V_1(\vp) = f_1 \exp{-\a\vp}$ \rf{V-1-def})
with running ``inflaton''-dependent gauge coupling constants:
\be
f_{\rm eff}(\vp) = f_0 \bigl(1-4\eps\chi_2 \cU (\vp)\bigr) \;\; ,\;\;
\frac{1}{e^2_{\rm eff}(\vp)} = \frac{\chi_2}{e^2} \Bigl\lb 1 +
\eps e^2 f_0^2 \bigl(1-4\eps\chi_2 \cU (\vp)\bigr)\Bigr\rb
\lab{running-const}
\ee

Concluding this Section let us note that $f_{\rm eff}(\vp)$ measures the
strength of charge confinement. Indeed, as shown in Appendix B of
\ct{grav-bags}, for static spherically symmetric fields in a static
spherically symmetric metric the presence of the term 
$-\h f_{\rm eff}(\vp) \sqrt{-{\bar F}^2}$ will produce an effective
``Cornell''-type \ct{cornell} potential $V_{\rm eff} (L)$ between charged 
quantized fermions, $L$ being the distance between the latter:
\be
V_{\rm eff} (L) = - \frac{e^2_{\rm eff}(\vp)}{2\pi} \frac{1}{L} + 
e_{\rm eff}(\vp) f_{\rm eff}(\vp) \sqrt{2}\, L +
\bigl( L{\rm -independent} ~{\rm const} \bigr) \; ,
\lab{cornell-type}
\ee
\textsl{i.e.}, a linear confining plus a Coulomb piece.
Thus, we will consider below the dynamics of the Abelian nonlinear gauge
system as mimicking quantum chromodynamics.

%%%%%%%%%%%%%%%%%%%%%%%%%%%%%%%%%%%%%%%%%%%%%%%%%%%%%%%%%%%%%%%%%%%%%%%%%%%%%%%%%%
\section{Cosmological Implications}
\label{cosmolog}
%%%%%%%%%%%%%%%%%%%%%%%%%%%%%

\subsection{Flat Regions of the Total ``Inflaton'' Effective Potential}
\label{flat-regions}

The total matter effective potential \rf{U-total} possesses non-trivial
``inflaton''-dependent gauge field vacuum 
($\partder{}{{\bar F}^2} U_{\rm total}\bv_{{\bar F}^2_{\rm vac}} = 0$):
\be
\sqrt{-{\bar F}^2_{\rm vac}} = e^2_{\rm eff}(\vp) f_{\rm eff}(\vp) \; ,
\lab{F-vac}
\ee
Upon inserting \rf{F-vac} in \rf{U-total} we get the following total
effective ``inflaton'' potential $U_{\rm eff}(\vp)$ (taking into account
expressions \rf{cU-def}-\rf{running-const}):
\be
U_{\rm eff} (\vp) = \cU (\vp) + 
\frac{1}{4} e^2_{\rm eff}(\vp) f^2_{\rm eff} (\vp) =
\cU (\vp) + \frac{e^2 f_0^2 \( 1-4\eps\chi_2 \cU(\vp)\)^2}{
4\chi_2 \lb 1 + e^2\eps f_0^2 \( 1-4\eps\chi_2 \cU (\vp)\)\rb}
\lab{U-eff}
\ee

As clearly seen from the graphic representation of \rf{U-eff} on Fig.1, the 
total effective ``inflaton'' potential possesses the following reparkable
property -- two infinitely large flat regions corresponding to large
negative and large positive ``inflaton'' values (the sub-/super-scripts $(\pm)$
label belonging to the respective flat region):

%%%%%%%%%%%%%%%%%%%%%%%%%%%%%
\begin{figure}
\begin{center}
\includegraphics{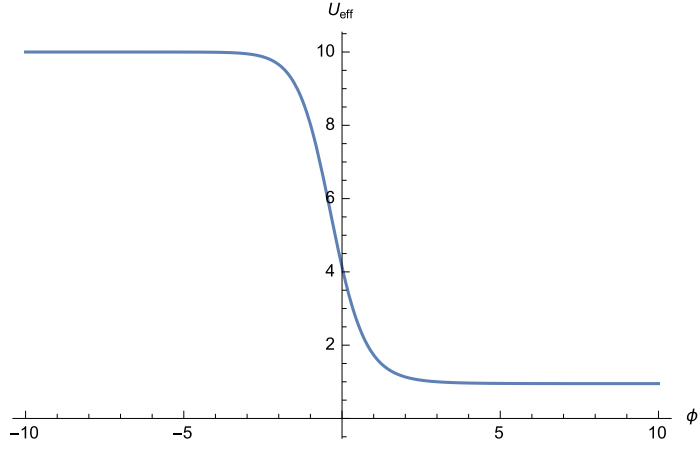}
\caption{Qualitative shape of the total effective ``inflaton'' potential 
$U_{\rm eff}(\vp)$ \rf{U-eff}.}
\end{center}
\end{figure}
%%%%%%%%%%%%%%%%%%%%%%%%%%%%%

\begin{itemize}
%%%%
\item
$(-)$ flat region -- for large negative $\vp$-values:
\br
\cU(\vp) \simeq \cU_{(-)} = \frac{1}{4\eps\chi_2} \quad ,\quad
U_{\rm eff} (\vp) \simeq U_{\rm eff}^{(-)} = \cU_{(-)} =
\frac{1}{4\eps\chi_2} \; ,
\lab{U-minus} \\
f_{\rm eff} (\vp)\simeq f_0^{(-)} = 0 \quad , \quad
e^2_{\rm eff}(\vp) \simeq e^2_{(-)} = \frac{e^2}{\chi_2} \; .
\lab{f-e-minus}
\er
The first relation in \rf{f-e-minus} implies according to \rf{cornell-type}
that there is \textbf{{\em no} charge confinement in the $(-)$ flat region}.
%%%%
\item
$(+)$ flat region -- for large positive $\vp$-values:
\br
\cU(\vp) \simeq \cU_{(+)} = 
\frac{M_1^2}{4 \llb \chi_2 (M_2 + \eps M_1^2) - 2M_0\rrb} \; ,
\lab{U-plus} \\
U_{\rm eff} (\vp) \simeq U_{\rm eff}^{(+)} = \frac{\eps\chi_2 M_1^2 
+ \eps e^2 f_0^2 (\chi_2 M_2 - 2 M_0)}{4\eps\chi_2 
\llb (\chi_2 M_2 - 2 M_0)(1+\eps e^2 f_0^2) +\eps\chi_2 M_1^2\rrb} \; ,
\lab{U-eff-plus} \\
f_{\rm eff} (\vp)\simeq f_0^{(+)} = f_0
\frac{(\chi_2 M_2 - 2 M_0)}{\chi_2 M_2 - 2 M_0 + \eps\chi_2 M_1^2}  \; ,
\lab{f-plus} \\
\frac{1}{e^2_{\rm eff}(\vp)} \simeq \frac{1}{e^2_{(+)}} =
\frac{\chi_2}{e^2} \Bigl\lb 1 + \frac{\eps e^2 f_0^2 (\chi_2 M_2 - 2 M_0)}{
\chi_2 M_2 - 2 M_0 + \eps\chi_2 M_1^2}  \Bigr\rb \; .
\lab{e-plus}
\er
Relation \rf{f-plus} implies according to \rf{cornell-type}
that \textbf{charge confinement {\em does take place}
in the $(+)$ flat region}. Also, there is a non-trivial rescaling of the
fundamental electric charge unit when passing from the $(-)$ flat
``inflaton'' region (second relation in \rf{f-e-minus})  
to the $(+)$ flat ``inflaton'' region \rf{e-plus}.
%%%%
\end{itemize}

Let us specifically point out that both the vacuum energy densities 
$U_{\rm eff}^{(\pm)}$ as well as the confinement/deconfinement phenomena in
both flat ``inflaton'' regions are entirely {\em dynamically} induced.

Associating appropriately the integration constants $M_{1,2},\, M_0$ and the 
coupling constants $f_0,\,\eps$ with the fundamental physical constants
$M_{EW}$ (electroweak scale), $M_{Pl}$ (Planck mass) and $M_{infl}$
(inflationary energy scale):
\be
M_1 \sim M_{EW}^4 \;\;, \;\; M_2 \sim M_{Pl}^4 \;\; ,\;\;
e^2 f_0^2 \sim \frac{M_1}{M_2} \;\; ,\;\; \chi_2 \eps \sim M_{infl}^{-4} \;,
\lab{scales}
\ee
allow us to identify -- through their respective vacuum energy densities 
$U_{\rm eff}^{(-)} \sim 10^{-8} M_{Pl}^4 \sim M_{infl}^4$ \rf{U-eff-plus} and 
$U_{\rm eff}^{(+)} \sim 10^{-122} M_{Pl}^4$ \rf{U-minus} --  the $(-)$ flat 
``inflaton'' region as describing the ``early'' Universe, whereas the 
$(+)$ flat ``inflaton'' region will be describing the ``late'' Universe in
accordance with the PLANCK collaboration data \ct{Planck}.
%%%%%%%%%%%%%%%%%%%%%%%%%%%%%

\subsection{FLRW Reduction and Dark Energy/Dark Matter Unification}
\label{FLRW}

Let us now consider a reduction of the Einstein-frame gauge-matter action
\rf{L-eff} to the class of cosmological FLRW class of metrics \rf{FLRW-0}
(here for simplicity we will take $K=0$ -- flat spacial sections):
\be
ds^2 = {\bar g}_{\m\n} dx^\m dx^\n = - N^2 (t) dt^2 + a^2(t) d{\vec x}.d{\vec x} 
\; ,
\lab{FLRW}
\ee
where now ${\bar X}=\h \vpdot^2\!\!\!(t)$, ${\wti Y} = \h \utdot^2\!\!\!(t)$, and 
$\sqrt{-{\bar F}^2} = \Adot\!\!(t)$. The resulting full FLRW Einstein-frame action
reads (in the gauge $N(t)=1$):
\be
S_{\rm FLRW} = \int dt \Bigl\lb - 6 a \adot^2 + a^3 L_{\rm FLRW}\Bigr\rb
\lab{S-FLRW}
\ee
with the FLRW matter-gauge Lagrangian, or the pressure $p$:
\br
p \equiv L_{\rm FLRW} = - \frac{f_0}{2} \Bigl\lb 1+\eps\chi_2 \Bigl(
\vpdot^2 - \utdot^2 (V_1 (\vp) + M_1)\Bigr)\Bigr\rb \Adot
\nonu \\
+ \frac{\chi_2}{4e^2} (1+\eps e^2 f_0^2) \Adot^2 + \frac{\vpdot^2}{2}
+\eps\chi_2 \frac{\vpdot^4}{4} - \frac{\utdot^2}{2}\bigl( V_1(\vp)+M_1\bigr)
(1+\eps\chi_2 \vpdot^2)
\nonu \\
+ \frac{\utdot^4}{4}\Bigl\lb 
\chi_2 M_2 - 2 M_0 + \eps\chi_2 (V_1(\vp)+ M_1)^2 \Bigr\rb \; .
\lab{L-eff-FLRW}
\er
The canonical FLRW Hamiltonian corresonding to $L_{\rm FLRW}\equiv p$ 
\rf{L-eff-FLRW}, or the energy density $\rho$ is given by:
\be
\rho = \frac{\pi_A}{a^3} + \frac{p_{\vp}}{a^3} + \frac{p_{\wti u}}{a^3}
- L_{\rm FLRW} \; ,
\lab{rho-FLRW}
\ee
where the canonical momenta are defined via:
\be
\frac{\pi_A}{a^3} = \partder{ L_{\rm FLRW}}{\Adot} \quad ,\quad
\frac{p_\vp}{a^3} = \partder{ L_{\rm FLRW}}{\vpdot} \quad ,\quad
\frac{p_{\wti u}}{a^3} = \partder{ L_{\rm FLRW}}{\utdot} \; .
\lab{can-momenta}
\ee
From the first Eq.\rf{can-momenta} we find:
\be
\Adot = \frac{2e^2}{\chi_2 (1+\eps e^2 f_0^2)}\Bigl\{\frac{\pi_A}{a^3} +
\frac{f_0}{2} \Bigl\lb 1+\eps\chi_2 \Bigl(
\vpdot^2 - \utdot^2 (V_1(\vp) + M_1)\Bigr)\Bigr\rb\Bigr\} \; ,
\lab{Adot-eq}
\ee
whereas the second and the third Eqs.\rf{can-momenta} are a system of two
mixed cubic algebraic equations w.r.t. the velocities $\vpdot$ and
$\utdot$:
\br
\frac{p_\vp}{a^3} = \frac{\eps\chi_2\,\vpdot^3}{1+\eps e^2 f_0^2}
+ \frac{\vpdot}{1+\eps e^2 f_0^2} \Bigl\lb 1 - 2\eps e^2 f_0 \frac{\pi_A}{a^3}
- \eps\chi_2 (V_1(\vp)+M_1) \utdot^2 \Bigr\rb \; ,
\lab{cubic-vpdot} \\
\frac{p_{\wti u}}{a^3} = \utdot^3 \Bigl\lb \chi_2 M_2 - 2 M_0 
+ \frac{\eps\chi_2}{1+\eps e^2 f_0^2} (V_1(\vp)+ M_1)^2 \Bigr\rb 
\nonu \\
- \utdot \frac{(V_1(\vp)+ M_1)}{1+\eps e^2 f_0^2}\Bigl(1 
- 2\eps e^2 f_0 \frac{\pi_A}{a^3} - \eps\chi_2 \vpdot^2\Bigr) \; .
\lab{cubic-utdot}
\er

\bigskip
\textbf{(i) In the $(+)$ flat region of the total effective
``inflaton'' potential, i.e. in the ``late'' Universe} where $\vp$ is large
positive, negleting $V_1(\vp) =  f_1 e^{-\a\vp}$ in 
\rf{cubic-vpdot}-\rf{cubic-utdot}, we obtain approximate (up to higher
powers of $\frac{1}{a^3}$) solutions for the velocities $\vpdot$ and
$\utdot$ as functions of the canonical momenta:
\br
\vpdot = \frac{p_\vp}{a^3}\,(1+\eps e^2 f_0^2)\,
\frac{(\chi_2 M_2 -2 M_0)(1+\eps e^2 f_0^2) + \eps\chi_2 M_1^2}{
(\chi_2 M_2 -2 M_0)(1+\eps e^2 f_0^2)} + \mathrm{O}(a^{-6}) \; ,
\lab{vpdot-plus} \\
\utdot = \Bigl\lb\frac{M_1}{(\chi_2 M_2 -2 M_0)(1+\eps e^2 f_0^2) + \eps\chi_2
M_1^2}\Bigr\rb^{\h} + \h (1+\eps e^2 f_0^2) \times
\nonu \\
% \h (1+\eps e^2 f_0^2)\,
\times \Bigl\lb M_1 \Bigl((\chi_2 M_2 -2 M_0)(1+\eps e^2 f_0^2) 
+ \eps\chi_2 M_1^2\Bigr)\Bigr\rb^{-\h}\,\frac{p_{\wti u}}{a^3} 
+ \mathrm{O}(a^{-6}) \; .
\lab{utdot-plus}
\er
Inserting \rf{vpdot-plus}-\rf{utdot-plus} into \rf{L-eff-FLRW}-\rf{rho-FLRW}
we obtain for the pressure and energy density in the ``late'' Universe:
\br
p_{(+)} = - U^{(+)}_{\rm eff} + \mathrm{O}(a^{-6}) \; ,
\lab{pressure-plus} \\
\rho_{(+)} = U^{(+)}_{\rm eff} + 
\Bigl\{ p_{\wti u}\,\Bigl\lb\frac{M_1}{(\chi_2 M_2 -2 M_0)(1+\eps e^2 f_0^2)
+ \eps\chi_2 M_1^2}\Bigr\rb^{\h} 
\nonu \\
+ \pi_A\,\frac{e^2 f_0}{\chi_2 (1+\eps e^2 f_0^2)}
\,\frac{(\chi_2 M_2 -2 M_0)(1+\eps e^2 f_0^2) - \eps\chi_2 M_1^2}{
(\chi_2 M_2 -2 M_0)(1+\eps e^2 f_0^2) + \eps\chi_2 M_1^2}\Bigr\} \,
\frac{1}{a^3} + \mathrm{O}(a^{-6}) \; ,
\lab{rho-plus}
\er
where $U^{(+)}_{\rm eff}$ is the dynamically induced vacuum energy density in
the ``late'' Universe \rf{U-eff-plus}. Note that there is {\em no} 
$\mathrm{O}(a^{-3})$ in the expression for the pressure \rf{pressure-plus},
while the leading terms in \rf{pressure-plus}-\rf{rho-plus} are with
opposite signs. Thus, comparing with \rf{p-rho-FLRW} we can identify the
expressions \rf{pressure-plus}-\rf{rho-plus} as describing a unification in
the ``late'' Universe of the dark energy 
($\rho^{\rm DE}_{(+)} = - p^{\rm DE}_{(+)} = U^{(+)}_{\rm eff}$) and dust dark
matter ($p^{\rm DM}_{(+)} = 0$, $\rho^{\rm DM}_{(+)} = \mathrm{O}(a^{-3})$ 
term in \rf{rho-plus}) contributions..

\bigskip

\textbf{(ii) In the $(-)$ flat region of the total effective
``inflaton'' potential, i.e. in the ``early'' Universe} where $\vp$ is large
negative, the approximate solutions for $\vpdot$ and $\utdot$ as functions
of the canonical momenta -- counterparts of Eqs.\rf{vpdot-plus}-\rf{utdot-plus} 
-- read:
\br
\vpdot = 
\Bigl\lb\frac{(1+\eps e^2 f_0^2) p_\vp}{2\eps\chi_2}\Bigr\rb^{1/3}\, a^{-1}
+ \mathrm{O}(a^{-2}) \; ,
\lab{vpdot-minus} \\
\utdot = \frac{1}{\sqrt{\eps\chi_2 f_1}} - \h \sqrt{\frac{\eps\chi_2}{f_1}} 
\Bigl\lb\frac{(1+\eps e^2 f_0^2) p_\vp}{2\eps\chi_2}\Bigr\rb^{2/3}\, a^{-2} 
+ \mathrm{O}(a^{-3}) \; .
\lab{utdot-minus}
\er
Inserting \rf{vpdot-minus}-\rf{utdot-minus} into \rf{L-eff-FLRW}-\rf{rho-FLRW}
we obtain for the pressure and energy density in the ``early'' Universe:
\br
p_{(-)} = - U^{(-)}_{\rm eff} + \mathrm{O}(a^{-4}) \; ,
\lab{pressure-minus} \\
\rho_{(+)} = U^{(-)}_{\rm eff} + 
\frac{p_{\wti u}e^{\h\vp}}{\sqrt{\eps\chi_2 f_1}}\, a^{-3} +
\mathrm{O}(a^{-4}) \; ,
\lab{rho-minus}
\er
where $U^{(-)}_{\rm eff}$ is the dynamically induced vacuum energy density in
the ``early'' Universe \rf{U-minus}. Once again we see that there is {\em no} 
$\mathrm{O}(a^{-3})$ in the expression for the pressure \rf{pressure-minus}
as in \rf{pressure-plus},
while the leading terms in \rf{pressure-minus}-\rf{rho-minus} are with
opposite signs. Thus, comparing with \rf{p-rho-FLRW} we can again identify the
expressions \rf{pressure-minus}-\rf{rho-minus} as describing a unification in
the ``early'' Universe of the dark energy 
($\rho^{\rm DE}_{(-)} = - p^{\rm DE}_{(-)} = U^{(-)}_{\rm eff}$) and dust dark
matter ($p^{\rm DM}_{(-)} = 0$, $\rho^{\rm DM}_{(-)} = \mathrm{O}(a^{-3})$
term in \rf{rho-minus}) contributions.

%%%%%%%%%%%%%%%%%%%%%%%%%%%%%%%%%%%%%%%%%%%%%%%%%%%%%%%%%%%%%%%%%%%%%%%%%%%%%%%%%%
\section{Conclusions}
\label{conclude}

We have demonstrated above that the method of non-Riemanninan spacetime
volume-forms (metric-independent volume elements), combined with other 
ingredients such as non-canonical gravity interactions with ``inflaton'' and 
``darkon'' scalar fields and special type of non-linear gauge fields, 
provides a plausible description of various basic features of cosmological
evolution:

(a) Unified description of dark energy and dark matter
% as manifestation of a single material entity 
-- through the impact of the dynamics of the  non-canonical scalar ``darkon'' field;

(b) Natural explanation of the huge difference between the vacuum energy
scales in the ``early'' and ``late'' Universe -- thanks to the % dynamical
generation of dimensionful integration constants as an impact of the
dynamics of the non-Riemannian volume-form fields;;

(c) Natural explanation of the absence of charge confinement in the
``early'' Universe epoch, and the appearance of QCD-like confinement in the
``late'' Universe - through an ``inflaton''-running coupling constants in the
physical Einstein-frame theory.

Further physically relevant phenomena adequately described using the method
of non-Riemanninan spacetime volume-forms are:

(d) A novel mechanism for the supersymmetric Brout-Englert-Higgs effect, namely, 
the appearance of a dynamically generated cosmological constant triggering 
spontaneous supersymmetry breaking and mass generation for the gravitino 
\ct{susyssb-1,susyssb-2}. 

(e) Gravity-assisted spontaneous symmetry breaking of electroweak gauge
symmetry -- when adding interactions with the bosonic fields of the
electroweak sector of the standard model of elementary particles gravity
triggers the generation of a Higgs-like spontaneous symmetry breaking effective 
potential in the ``late'' Universe, whereas in the ``early'' Universe 
gravity keeps the Higgs-like scalar isodoublet massless, \textsl{i.e.}, 
no spontaneous electroweak breaking in the ``early'' Universe \ct{grf-essay}.

Finally, let us mention some other modifications of the method of
non-Riemannian volume elements -- the so called gravity models with dynamical 
spacetime \ct{eduardo-dynamical-time}, which were further developed into
models of interacting diffusive unified dark energy and dark matter 
(\ct{benisty-eduardo} and references therein).

%%%%%%%%%%%%%%%%%%%%%%%%%%%%%%%%%%%%%%%%%%%%%%%%%%%%%%%%%%%%%%%%%%%%%%%%%%%%%%%%%%
\section*{Acknowledgments} 
E.G. and E.N. are sincerely grateful to Prof. Branko Dragovich and the organizers
of the {\em Ninth Meeting in Modern Mathematical Physics} in Belgrade for cordial 
hospitality. 
E.G., E.N. and S.P. gratefully acknowledge support of our collaboration through 
the academic exchange agreement between the Ben-Gurion University in Beer-Sheva,
Israel, and the Bulgarian Academy of Sciences. 
E.N. and E.G. have received partial support from European COST actions
MP-1405 and CA-16104, and from CA-15117 and CA-16104, respectively.
E.N. and S.P. are also thankful to Bulgarian National Science Fund for
support via research grant DN-18/1. 

%%%%%%%%%%%%%%%%%%%%%%%%%%%%%%%%%%%%%%%%%%%%%%%%%%%%%%%%%%%%%%%%%%%%%%%%%%%%%%%%%%
%%%%%%%%%%%%%%%%%%%%%%%%%%%%%%%%%%%%%%%%%%%%%%%%%%%%%%%%%%%%%%%%%%%%%%%%%%%%%%%%%%

\end{document}